\documentclass[longauth,letter]{aa}
\usepackage{graphicx}
\usepackage{txfonts}
\usepackage{lipsum}
\usepackage{subfigure}
\usepackage{subcaption}         
\usepackage{lscape}             
\usepackage{placeins}           
\usepackage{xcolor}
\newcommand{\review}[1]{{#1}}

\usepackage{hyperref}

\begin{document}

   \title{A faint dust ring beyond the compact disk of Sz~76}

   \author{G. Ricciardi\inst{\ref{instESO}}
   \corrauth{giulia.ricciardi@eso.org} 
        \and A. Miotello\inst{\ref{instESO}}
        \and K. Mauco\inst{\ref{UNAM}}
        \and C. F. Manara\inst{\ref{instESO}}
        \and S. Facchini\inst{\ref{UniMI}}
        \and \'A. K\'osp\'al\inst{\ref{csfk}}
        \and J. Alcal\'a\inst{\ref{instNA}}
        \and J. Eisl\"offel\inst{\ref{instTLS}}
        \and C.C. Espaillat\inst{\ref{cce}}
        \and M. Vioque\inst{\ref{instESO}}
        \and J.~P.~Williams\inst{\ref{instIfA}}
        \and Zhen~Guo\inst{\ref{instValP},\ref{instCAS}}
        }

   \institute{European Southern Observatory, Karl-Schwarzschild-Strasse 2, 85748 Garching bei M\"unchen, Germany\label{instESO}
   \and Instituto de Astronomía, Universidad Nacional Autónoma de México, Km. 107 Carretera Tijuana-Ensenada, 22860 Ensenada, B.C., México\label{UNAM}
   \and Dipartimento di Fisica, Università degli Studi di Milano, Via Celoria 16, 20133 Milano, Italy\label{UniMI}
   \and INAF – Osservatorio Astronomico di Capodimonte, via Moiariello 16, 80131 Napoli, Italy\label{instNA}
   \and Th\"uringer Landessternwarte, Sternwarte 5, D-07778 Tautenburg, Germany\label{instTLS} \and Department of Astronomy, Boston University, 725 Commonwealth Avenue, Boston, MA 02215, USA\label{cce} \and Konkoly Observatory, HUN-REN Research Centre for Astronomy and Earth Sciences, MTA Centre of Excellence, Konkoly-Thege Miklós út 15–17, 1121 Budapest, Hungary\label{csfk}
   \and Institute for Astronomy, University of Hawai’i at M\~anoa, Honolulu, HI 96822, USA\label{instIfA}
\and 
Instituto de F{\'i}sica y Astronom{\'i}a, Universidad de Valpara{\'i}so, ave. Gran Breta{\~n}a, 1111, Casilla 5030, Valpara{\'i}so, Chile\label{instValP}
\and
Chinese Academy of Sciences South America Center for Astronomy (CASSACA), National Astronomical Observatories, CAS,
Beijing 100101, China\label{instCAS}
   }

   \date{\today}
 
  \abstract{We report new ALMA Band 7 high-resolution ($\sim$0.05$^{''}$) continuum and $^{12}\text{CO}\ (J=3\text{--}2)$ observations of the disk around the young low-mass star Sz 76. Visibility plane modeling have revealed a previously undetected, faint outer dust ring centered at $\sim$45 au surrounding the compact central disk. Including this low-surface-brightness feature increases the characteristic dust radius from $R_{\text{dust,68}} \approx 7$ au to $41$ au. Combined with an outer gas extent of $R_{\text{CO,68}} \approx 77$ au, the resulting gas-to-dust size ratio decreases from an extreme value of $\sim$11 to $\sim$2, setting Sz 76 in line with typical protoplanetary disks and contradicting a scenario of unimpeded inward radial drift. These results demonstrate that spatial filtering or limited surface-brightness sensitivity can severely underestimate disk sizes, highlighting the necessity of multi-scale observations to properly identify dust traps and accurately constrain disk evolution.}


   \maketitle
   \nolinenumbers

\section{Introduction}
The high incidence of compact protoplanetary disks ($R_{\rm dust}<20$ au) in millimeter continuum surveys conducted with ALMA \citep[see][as a review]{Manara2023} is broadly consistent with theoretical expectations from radial drift, which predicts that dust grains grow into pebbles and migrate toward local gas pressure maxima, leading to an inward redistribution of the solid material \citep[][]{Weidenschilling1997}. In disks with smooth pressure profiles, this process concentrates dust in the inner regions, producing continuum emission that is significantly more compact than the underlying gas disk. High-angular-resolution ALMA observations have revealed that many bright and large disks host rings, gaps, and other radial substructures \citep[e.g.,][as reviews]{Andrews2020,Bae2023} that have commonly been interpreted as manifestations of local pressure maxima that efficiently trap drifting pebbles at large radii \citep{Pinilla2012,Dullemond2018}; in addition, some also exhibit faint continuum emission at large radii that had previously gone undetected \citep{Ilee2022, Das24, Curone25}. On the other hand, our understanding of compact disks, which constitutes a large fraction of the disk demographic, remains limited \citep{Miotello2021}. It is unclear whether compact disks host analogous substructures on smaller spatial scales or whether their dust distributions are primarily shaped by unimpeded radial drift. Radial drift models predict that millimeter continuum emission should be significantly less extended than molecular gas tracers, particularly CO \citep{Facchini2017,Trapman19}. Observational surveys of bright disks generally report gas-to-dust size ratios between $\sim$1.5 and 3 \citep{Barenfeld2017,Ansdell2018}. While some systems require radial drift to account for the observed discrepancy between gas and dust extents \citep[e.g.,][]{Panic2009,Andrews2012,Sanchis2021}, in most cases, the difference can be explained by optical depth effects, since CO emission remains detectable at lower surface densities than the dust continuum \citep{Dutrey1998,GuilloteauDutrey1998,Facchini2017}. ALMA provides sufficiently sensitive observations of both dust and gas emission in compact disks to test whether these systems are simply scaled-down versions of their larger counterparts or instead represent disks in which radial drift has dramatically reduced the extent of the dust component. 

Extreme dust concentration was reported for CX Tau, where the large disparity between the CO and millimeter continuum radii was interpreted as a clear signature of radial drift \citep{Facchini2019}. However, the recent survey of 18 faint and compact disks in Lupus by \cite{Ricciardi26} reached a different conclusion: all targets exhibit $R_{\rm gas}/R_{\rm dust} < 3$, with no compelling evidence for strong radial drift. In this context, CX Tau seems to be an exception more than the rule.  The target of this Letter is the disk around the protostar Sz~76 ($M_{\star}=$0.19 M$_{\odot}$, $\dot{M}_{\rm acc}=7.8 \cdot 10^{-10}~ M_{\odot}{\rm yr}^{-1}$) in the young Lupus~I cloud \citep{Alcala2017,Manara2023}. This disk was among the faintest, apparently compact disks, of the ALMA survey of disks in Lupus \citep{Ansdell2018}. The CO radius of this disk was measured by \cite{Sanchis2021} and compared to the dust extension to find a relatively compact disk. Recently \cite{GuerraAlvarado25} reported very high-angular resolution ($\sim$0.03$^{''}$) observations hinting at the presence of a cavity in the inner region of a compact continuum emission.

In this Letter, we present ALMA Band 7 observations of Sz~76 at an angular resolution of $\sim 0.05''$, providing new insights into the properties of its gas and dust components. Sz 76 is an intriguing system due to its historically ambiguous evolutionary classification. Despite retaining an infrared excess and detectable millimeter continuum emission, it has been classified in the literature as a weak-line T Tauri star (WTTS) or Class III source \citep{Wahhaj10, LopezMartin11, Alcala2017}. Thus, Sz 76 is an interesting system for investigating the connection between disk dispersal and the radial evolution of solids, as it shows relatively weak accretion activity alongside a still detectable circumstellar disk. Section~\ref{sec2} describes the observations, Sect. \ref{sec3} presents the analysis and results, and Sect.~\ref{sec4} places these results within the broader context of disk evolution.

\section{Observations} \label{sec2}

\begin{figure}[htb!]
    \centering
    \includegraphics[width=0.7\columnwidth]{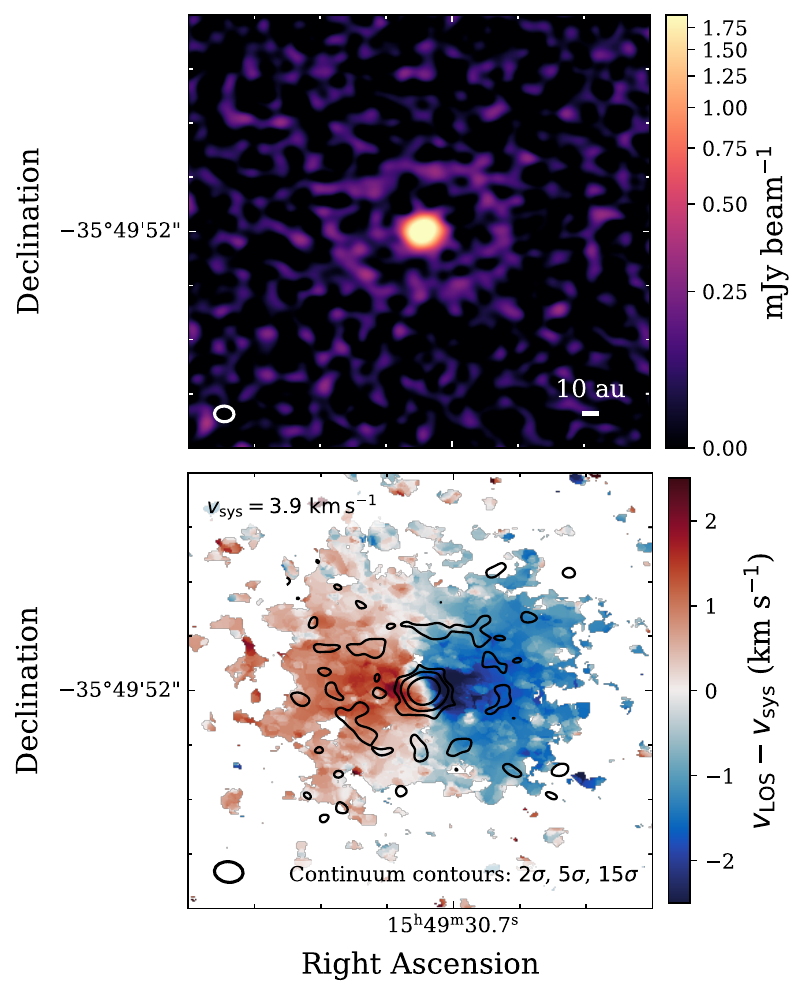}
    \vspace{-0.4cm}
    \caption{Band 7 continuum image (top), and intensity-weighted velocity (moment 1) map of the $^{12}$CO emission (bottom).}
    \label{images}
\end{figure}

The data presented in this Letter were obtained as part of the ALMA program $\#$2023.1.01276.S (PI: K. Mauco). The disk around Sz 76 was targeted in  H$^{13}$CN ($J=4-3$), $^{12}$CO ($J=3-2$), and HCN ($J=4-3$) line emission, and with two broad continuum windows. In this Letter, we focus exclusively on the continuum emission and the $^{12}$CO ($J=3-2$) line. For the continuum, we combined all line-free channels, resulting in a total effective bandwidth of 4.453 GHz, while the CO data were analyzed at a spectral resolution of 0.106 km s$^{-1}$ before any subsequent spectral averaging. The details on the data calibration and imaging are reported in Appendix \ref{App_data}. The continuum image and the intensity-weighted velocity $^{12}$CO emission map are presented in Fig.~\ref{images}. 

\begin{figure*}[htb!]
    \centering
    \begin{minipage}[c]{0.65\textwidth}
        \centering
        \includegraphics[width=\linewidth]{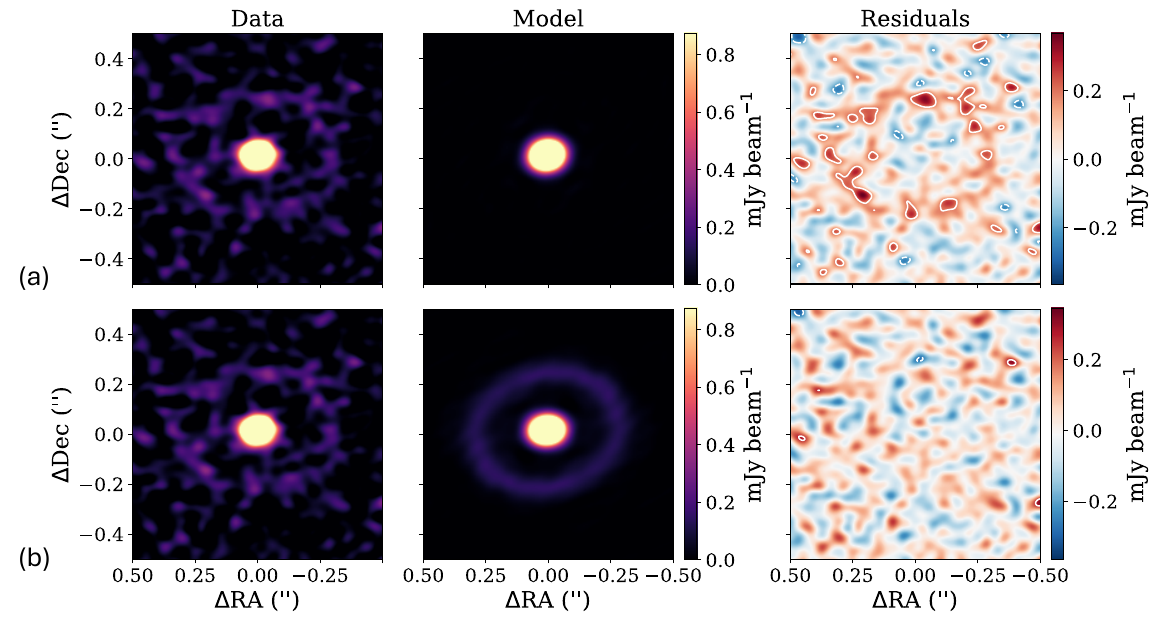}
    \end{minipage}
    \hfill
    \begin{minipage}[c]{0.25\textwidth}
        \caption{From left to right: Observed continuum image, best-fit model by Galario, and residual map obtained after subtracting the model from the observed visibilities. The model image and residuals were produced using the same imaging parameters adopted for the observed data. Contours in the residual map are drawn at $\pm3\sigma$. Panels (a) show the single Gaussian model of the visibilities; (b) the Gaussian+ring profile.}
        \label{data_mod_res}
    \end{minipage}
\end{figure*}

\section{Results} \label{sec3}
\subsection{Dust modeling}
To characterize the dust continuum emission, we modeled the ALMA visibilities directly in the uv-plane using the Python package GALARIO \citep{Tazzari18}, which computes synthetic interferometric visibilities from an analytical surface-brightness model through a Fourier transform. The model parameters were explored using a Markov chain Monte Carlo (MCMC) approach by minimizing the $\chi^2$ between the observed and synthetic visibilities. Modeling the interferometric data directly in the uv-plane avoids biases introduced by image deconvolution and provides robust estimates of the intrinsic disk properties. The modeling was primarily aimed at determining the global disk geometry, such as the position angle (PA), inclination ($i$), and offsets in RA and Dec with respect to the phase center, together with the total continuum flux and the radius enclosing 68\% of the dust emission ($R_{68,\rm dust}$). To compute the dust radial profile, we considered two different profiles: a single Gaussian component representing the bulk of the dust emission (see Fig.\ref{data_mod_res}, panel a); a centrally peaked Gaussian component and an additional Gaussian ring (see Fig.~\ref{data_mod_res}, panel b). The residuals reveal that the single-Gaussian model leaves significant ring-shaped residual emission, suggesting that it does not accurately reproduce the observed brightness distribution. In contrast, the Gaussian+Ring model removes these residuals completely, providing a substantially better representation of the continuum emission. The presence of the faint ring is further supported by the deprojected visibility profile (Fig.~\ref{galario_plots_1ring}), where the ring model reproduces the oscillatory behavior (uv-distance $\simeq$500 k$\lambda$) of the visibilities that cannot be accounted for by a single Gaussian component (Fig.~\ref{galario_plots_only_gauss}). The best-fit Gaussian+ring model returns the following best-fit parameters: position angle PA = 104.82$^{+4.02}_{-4.00}$ deg, inclination $i$ = 36.87$^{+2.02}_{-2.19}$ deg, total flux F = 14.73$^{+1.02}_{-0.46}$ mJy, dust radius $R_{68,\rm dust}$ = 40.98$^{+2.8}_{-2.29}$ au, ring location $R_{\rm ring}$ = 45.36 $^{+1.56}_{-1.56}$ au, and ring width $\sigma_{\rm ring}$ = 4.69 $^{+1.56}_{-1.56}$ au at a distance of 156.4 pc.

\subsection{CO emission analysis}
The $^{12}$CO ($J=3-2$) emission is detected beyond $>40\sigma$ (Fig.~\ref{images}), revealing a gaseous component that extends beyond the faint continuum ring. To characterize the molecular gas, we used the Python package \texttt{GoFish} \citep{Teague19}, which provides specific tools for the analysis of interferometric spectral-line observations of protoplanetary disks. The integrated spectrum was obtained considering the PA, $i$, and disk center offsets derived from the continuum modeling, allowing the emission to be extracted from the disk reference frame. The total line flux derived is $F_{\rm CO}$ = 3.62$^{+0.24}_{-0.25}$ Jy km s$^{-1}$. To estimate the characteristic size of the gaseous disk, we applied the curve-of-growth method to the moment 0th map. A sequence of concentric elliptical apertures, centered on the continuum position and with PA and $i$ derived from the dust modeling, were used to measure the cumulative integrated flux as a function of aperture size \citep[e.g.,][]{Ricciardi26, Ansdell2018}. The cumulative flux increases with aperture until reaching a plateau, which provides an estimate of the total integrated emission. Here, we define the characteristic gas radius, $R_{68,\rm gas}$, as the disk extent enclosing 68\% of the total integrated flux. Specifically, $R_{68,\rm CO}$ = 77.2$^{+4.3}_{-4.9}$ au at 156.4 pc, showing that the molecular gas extends to nearly twice the radius measured for the dust continuum.

\subsection{Continuum intensity profile and comparison with the literature} \label{comparison}

\begin{figure}[htb!]
    \centering
    \includegraphics[width=0.8\columnwidth]{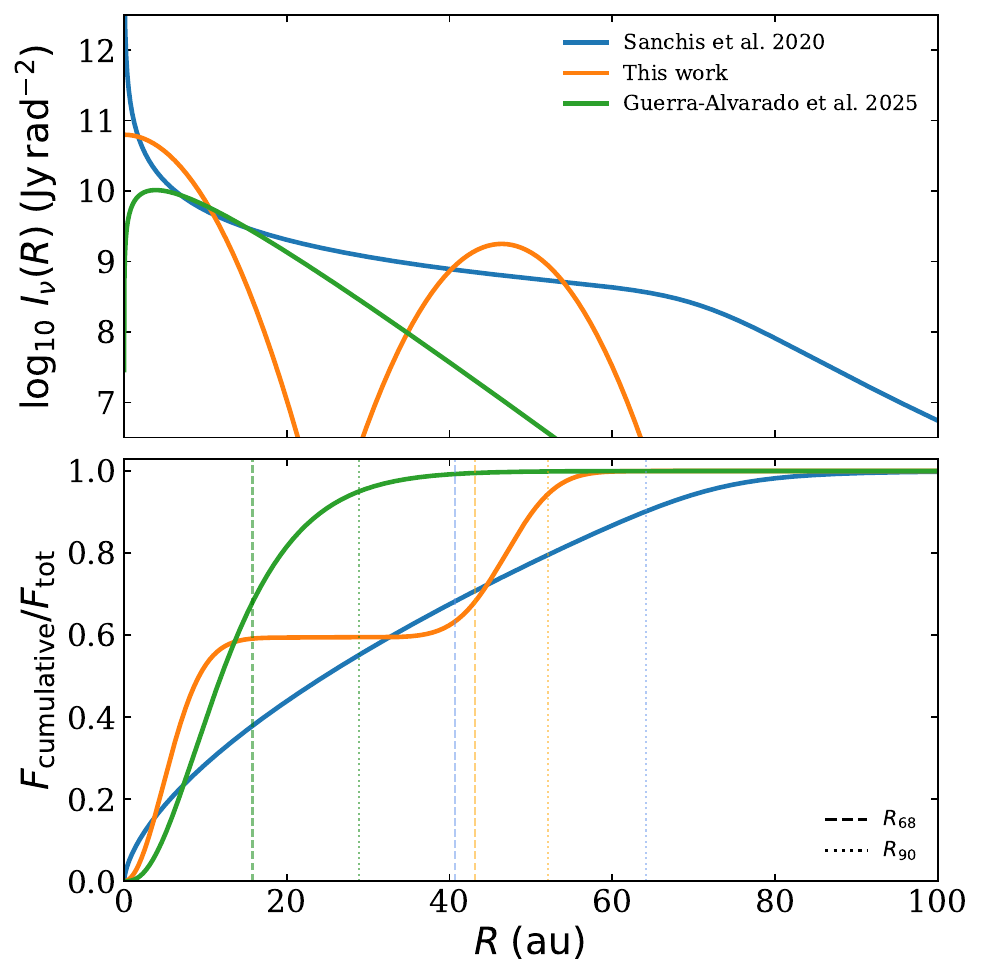}
    \vspace{-0.4cm}
    \caption{Sz 76 radial continuum intensity profiles inferred from different  visibility-domain models, adopting a common distance of 156.4 pc. \textit{Top:} Azimuthally averaged radial intensity profiles obtained from the Nuker model of \cite{Sanchis20}, the central Gaussian plus ring model presented in this work, and the compact-ring model adopted by \cite{GuerraAlvarado25}. \textit{Bottom:} Corresponding normalized cumulative flux profiles. Vertical dashed and dotted lines indicate $R_{68}$ and $R_{90}$, with colors matching the corresponding intensity profiles.}
    \label{IntensityProfile}
\end{figure}

Sz 76 was recently observed at very high angular resolution ($\sim0.03^{''}$) by \cite{GuerraAlvarado25} in ALMA Band 6. These observations have a synthesized beam of $0.026''\times0.025''$, which is approximately half the size of that in our Band 7 data, an rms of 0.019 mJy beam$^{-1}$ and a comparable surface-brightness sensitivity. The high angular resolution and, possibly, the lower frequency of their observations, made it possible for the innermost dust distribution to be resolved and a central depletion to be identified on scales of $\sim5$ au. However, their observations lacked short-baseline coverage, recovering only a fraction of the continuum flux measured at lower angular resolution. This suggests that a significant fraction of the extended emission was spatially filtered out. Thus, while their observations provide a more detailed view of the inner disk, they are insensitive to the faint emission at large radii revealed by our data. As shown in Fig.~\ref{IntensityProfile}, the best fit radial profile inferred by \cite{GuerraAlvarado25} shown in green is strongly concentrated towards the central disk - with an inner cavity - and declines rapidly with radius. Accordingly, its cumulative flux profile reaches the radii enclosing 68\% and 90\% of the total continuum flux of the model at substantially smaller radii ($R_{68,\rm dust}=15.78$ au and $R_{90,\rm dust}=24.27$ au) than in our model, shown in orange. Instead, our observations recover an additional, low-surface-brightness ring at $\sim45$ au, shifting the characteristic radii outward ($R_{68,\rm dust}=40.98$ au and $R_{90,\rm dust}=52.07$ au) and demonstrating how spatial filtering of faint extended emission can lead to a substantially underestimated disk size. Interestingly, \citet{Vioque26} recently identified a significant Gaia proper-motion anomaly for Sz 76 ($|\Delta\mu|/\sigma_{|\Delta\mu|}=4.2$), compatible within the uncertainties with a planetary-mass companion ($M_{\rm p}<13,M_{\rm J}$). The $\sim4$ au cavity adopted in their analysis is consistent with tidal truncation by the inferred companion; the somewhat larger inner depletion suggested by the high-resolution continuum observations therefore provides an interesting constraint on this interpretation.\\

Sz 76 was also observed in ALMA Band 7 by \cite{Sanchis2021}, with a synthesized beam of $\sim0.34''\times0.30''$ and an rms sensitivity of $\sim0.3$ mJy beam$^{-1}$. Although the lower angular resolution prevents the outer emission from being spatially resolved as a distinct ring, the larger beam provides enhanced brightness-temperature sensitivity to low-surface-brightness emission. Indeed, although no outer ring was directly detected in their images, \cite{Sanchis2021} found that a Nuker profile provided a better description of the continuum visibilities than a simple Gaussian profile. Therefore, the extended tail naturally produced by the Nuker model already offered a hint to suggest the presence of faint emission beyond the apparently compact central disk. Fig.~\ref{IntensityProfile} shows that the Nuker profile (shown in blue) does not reproduce the ring-like morphology revealed by our higher-resolution observations, but it distributes a significant fraction of the continuum flux over large radii. As a result, the value of $R_{68,\rm dust}=40.66$ au inferred by \citealt{Sanchis2021} is remarkably similar to ours, while the smooth outer tail of the Nuker model extends to substantially larger radii than the localized ring detected in our observations, resulting in a larger $R_{90,\rm dust}=64.12$ au. \\
As Fig.~\ref{IntensityProfile} shows, the inferred size and morphology of Sz 76 depend critically on the combination of angular resolution, surface-brightness sensitivity, and sampled spatial scales. Very high angular resolution reveals the innermost cavity; however, without adequate short-baseline coverage, it misses the faint outer disk. High brightness sensitivity can provide evidence for extended emission even when it cannot be spatially resolved. Thus, only observations combining sufficient angular resolution, sensitivity, and short- and long-baseline coverage can recover both components and distinguish the outer emission as a ring.

\section{Discussion and conclusions} \label{sec4}

With the new ALMA data presented in this work, we have discovered a faint dusty ring at $\sim45$ au from the protostar Sz~76, beyond a compact disk of $\sim 7$ au. This ring, together with the inner cavity reported by \cite{GuerraAlvarado25}, imply that complex sub-structured dust distributions can be found also around very-low-mass protostars ($M_{\star}=$0.19 M$_{\odot}$) and can be signposts of planet formation (see Appendix~\ref{DBnet} for more detail on exploratory planet–disk interaction modeling of Sz76).

We have also shown that the inferred dust distribution of a protoplanetary disk significantly depends on the observational setup \citep[see also][]{Ilee2022, Das24}. Adequate surface-brightness sensitivity is required to detect faint outer emission and short baselines are necessary to recover its total flux, while long baselines are needed to resolve fine structures. The analysis on Sz 76 disk emission demonstrates the physical consequences of this observational bias. Once the faint ring at $\sim45$ au was recovered with our Galario modeling, the measured dust radius increase from $R_{68,\rm dust}=7.1$ au for the bright central component alone to $R_{68,\rm dust}=40.98$ au. The detection of this faint outer ring therefore changes not only our view of the disk morphology, but also its evolutionary interpretation: while a dust disk confined within $\leq10$ au would naturally suggest efficient, unimpeded inward migration of millimeter-sized grains, the presence of an outer dust substructure instead points toward dust trapping in a pressure maximum. 

\begin{figure}[]
    \centering
    \includegraphics[width=\columnwidth]{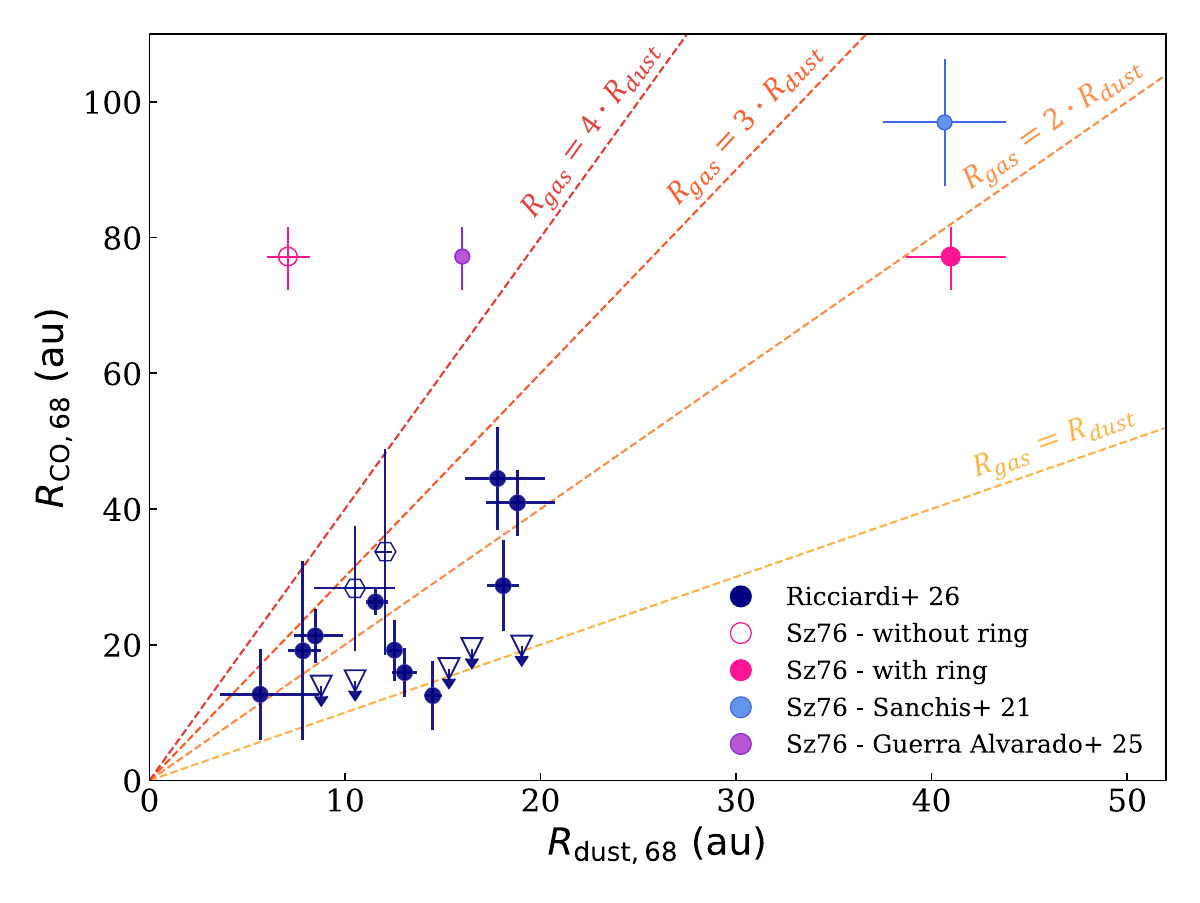}
    \vspace{-0.9cm}
    \caption{Comparison between dust and gas radii. The sample by \cite{Ricciardi26} is shown in blue. The empty and filled pink dots represent the Sz 76 extents from single-Gaussian and Gaussian+ring continuum models, respectively. The values from \cite{Sanchis2021} and \cite{GuerraAlvarado25} are shown in light blue and purple, respectively.}
    \label{dust_vs_gas}
\end{figure}

In this context, the simultaneous measurement of the gas and dust radii provides an important diagnostic of dust evolution. In particular, the gas-to-dust size ratio, $R_{\rm CO}/R_{\rm dust}$, can help constrain whether radial drift is a dominant process \citep{Facchini2017,Trapman19}. In our analysis, the $^{12}$CO ($J=3-2$) emission extends beyond the continuum, with $R_{68,\rm CO}=77.2$ au. We adopted $R_{68}$ for this comparison because the CO radius is measured from the image-plane curve of growth, whose outermost percentiles become increasingly sensitive to low-level noise and to the determination of the total-flux plateau \citep[see][as a review]{Miotello2023}. In particular, $R_{90}$ is controlled by the faintest emission at large radii and is therefore less robust for the present data. As shown in Fig.~\ref{dust_vs_gas}, considering only the bright inner dust component would yield to $R_{68,\rm gas}/R_{68,\rm dust}\simeq11$, far above the empirical threshold of $\sim4$ identified by \citet{Trapman19} as robust evidence for efficient radial drift. In a similar fashion, the $R_{68,\rm gas}/R_{68,\rm dust}$ ratio measured by \citep{GuerraAlvarado25} is above 4. Instead, once the faint outer ring is properly considered, the measured size ratio decreases to $\sim2$, placing Sz 76 among the bulk of the Lupus disk population, whose median value is $R_{68,\rm gas}/R_{68,\rm dust}\simeq2.5$ \citep{Sanchis2021}. This is fully consistent with the sample of radially compact disks analyzed by \citet{Ricciardi26}. As already pointed out in Sect. \ref{comparison}, the ratio obtained by the simplified modeling of \cite{Sanchis2021} is within the range of our result (see blue circle in Fig. \ref{dust_vs_gas}).

Sz 76 thus provides a cautionary example for the use of gas-to-dust size ratios as diagnostics of disk evolution. Insufficient sensitivity to low-surface-brightness continuum emission or an inadequate recovery of extended spatial scales can bias $R_{\rm dust}$ toward smaller values and artificially inflate $R_{\rm CO}/R_{\rm dust}$, mimicking the observational signature of efficient radial drift. More broadly, some disks currently classified as compact (and potentially interpreted as strongly affected by radial drift) could host faint outer dust structures that remain undetected. Establishing their true radial extent therefore requires observations that combine sufficient angular resolution and sensitivity with both short- and long-baseline coverage.

\begin{acknowledgements}
This paper makes use of the following ALMA data: ADS/JAO.ALMA\#2023.1.01276.S. ALMA is a partnership of ESO (representing its member states), NSF (USA) and NINS (Japan), together with NRC (Canada), NSTC and ASIAA (Taiwan), and KASI (Republic of Korea), in cooperation with the Republic of Chile. The Joint ALMA Observatory is operated by ESO, AUI/NRAO and NAOJ.
CFM and SF are funded by the European Union (ERC, WANDA, 101039452; ERC, UNVEIL, 101076613). Views and opinions expressed are however those of the author(s) only and do not necessarily reflect those of the European Union or the European Research Council Executive Agency. Neither the European Union nor the granting authority can be held responsible for them.
ZG is supported by FONDECYT Iniciación project 11260176. ZG is supported by the China-Chile Joint Research Fund (CCJRF No.2301) and the Chinese Academy of Sciences South America Center for Astronomy (CASSACA) Key Research Project E52H540301. 
\review{JMA acknowledges financial support from Large Grant INAF-2024 “Spectral Key features of Young stellar objects: Wind-Accretion LinKs Explored in the infraRed (SKYWALKER)” CUP C63C24001530005.
This work was also supported by the ADVANCED 149943 and 2024-1.2.8- T\'ET-IPARI-CN-2025-00036 grants, which have been implemented with the support provided by the Ministry of Culture and Innovation of Hungary from the National Research, Development and Innovation Fund, financed under the NKKP ADVANCED and 2024-1.2.8-T\'ET-IPARI-CN funding schemes.}
\end{acknowledgements}

\vspace{-0.5cm}
\bibliographystyle{aa}
\bibliography{biblio}

\begin{appendix}
\nolinenumbers
\onecolumn

\section{Details on the observations and data reduction} \label{App_data}

The data presented in this Letter were obtained as part of the ALMA program $\#$2023.1.01276.S (PI: K. Mauco). The source Sz 76 was observed on 31 May 2025 in Band 7 using 48 12-m antennas, with a total on-source integration time of 780.2 s (13.0 min). The observations were designed to achieve an angular resolution of $\sim$0.05$^{''}$, with a continuum sensitivity of $\sim$0.08 mJy beam$^{-1}$. The correlator was configured with five spectral windows, comprising two high-resolution windows centered on the H$^{13}$CN ($J=4-3$) and $^{12}$CO ($J=3-2$) transitions, one intermediate-bandwidth window centered on HCN ($J=4-3$), and two broad continuum windows covering the lower and upper sidebands. In this Letter, we focus exclusively on the continuum emission and the $^{12}$CO ($J=3-2$) line. For the continuum we combined all line-free channels, resulting in a total effective bandwidth of 4.453 GHz, while the CO data were analyzed at a spectral resolution of 0.106 km s$^{-1}$ before any subsequent spectral averaging. The data were calibrated using the CASA version 6.5.4-9 (Pipeline version 2023.1.0.125). To improve the image fidelity, three rounds of phase-only self-calibration were performed on the continuum visibilities using the CASA task \texttt{gaincal}, with solution intervals of solint=inf, 60 s, and 10 s, respectively. The derived calibration tables were sequentially applied to the visibility data leading to an improvement of $\sim 30\%$ in the signal-to-noise (S/N) ratio of the continuum emission. The self-calibration solutions were also applied to the $^{12}$CO data. The self-calibrated continuum visibilities were imaged with the CASA task \texttt{tclean} using Briggs weighting with a robust parameter of 0.5, and a multiscale deconvolution. Cleaning was performed down to a threshold of 0.1 mJy beam$^{-1}$ using the auto-multithresh masking algorithm. The $^{12}$CO ($J=3-2$) data were continuum-subtracted in the visibility domain and subsequently imaged. The data cube was produced using natural weighting to maximize surface-brightness sensitivity and a multiscale deconvolution. Cleaning was performed down to a threshold of 1 mJy beam$^{-1}$ using the auto-multithresh masking algorithm. The final continuum image and the intensity-weighted velocity $^{12}$CO emission map are presented in Fig.~\ref{images}. 

\section{Galario best-fit plots} \label{Galario_plots}

\begin{figure}[htb!]
\centering
\subfigure{
   \includegraphics[width=8cm]{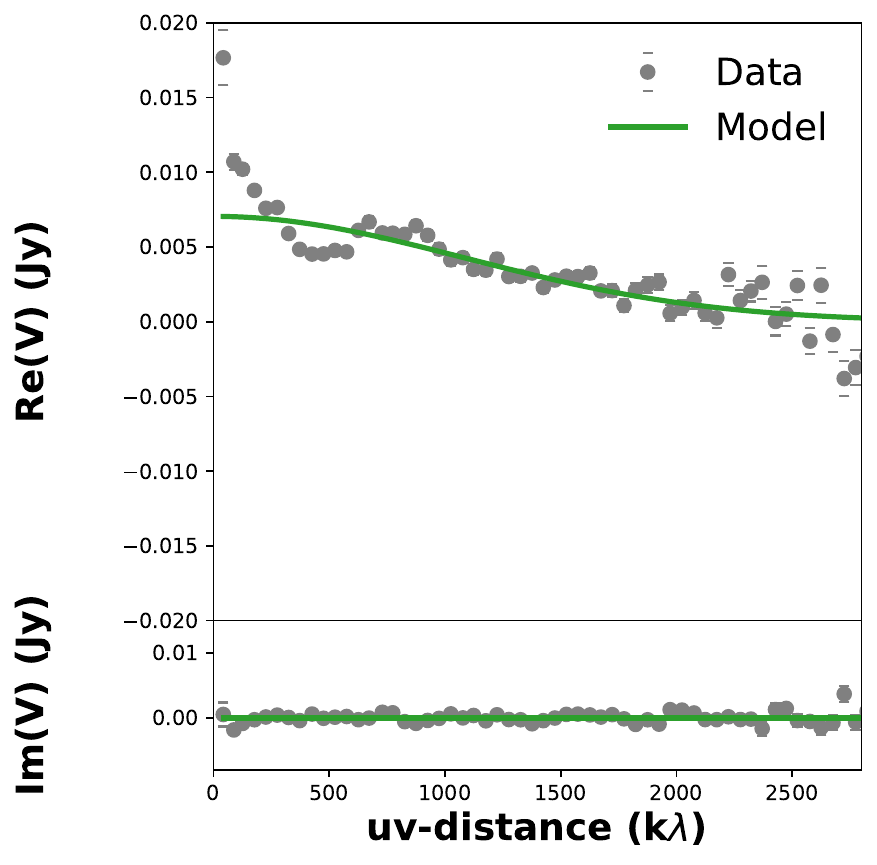}
}
\subfigure{
   \includegraphics[width=8cm]{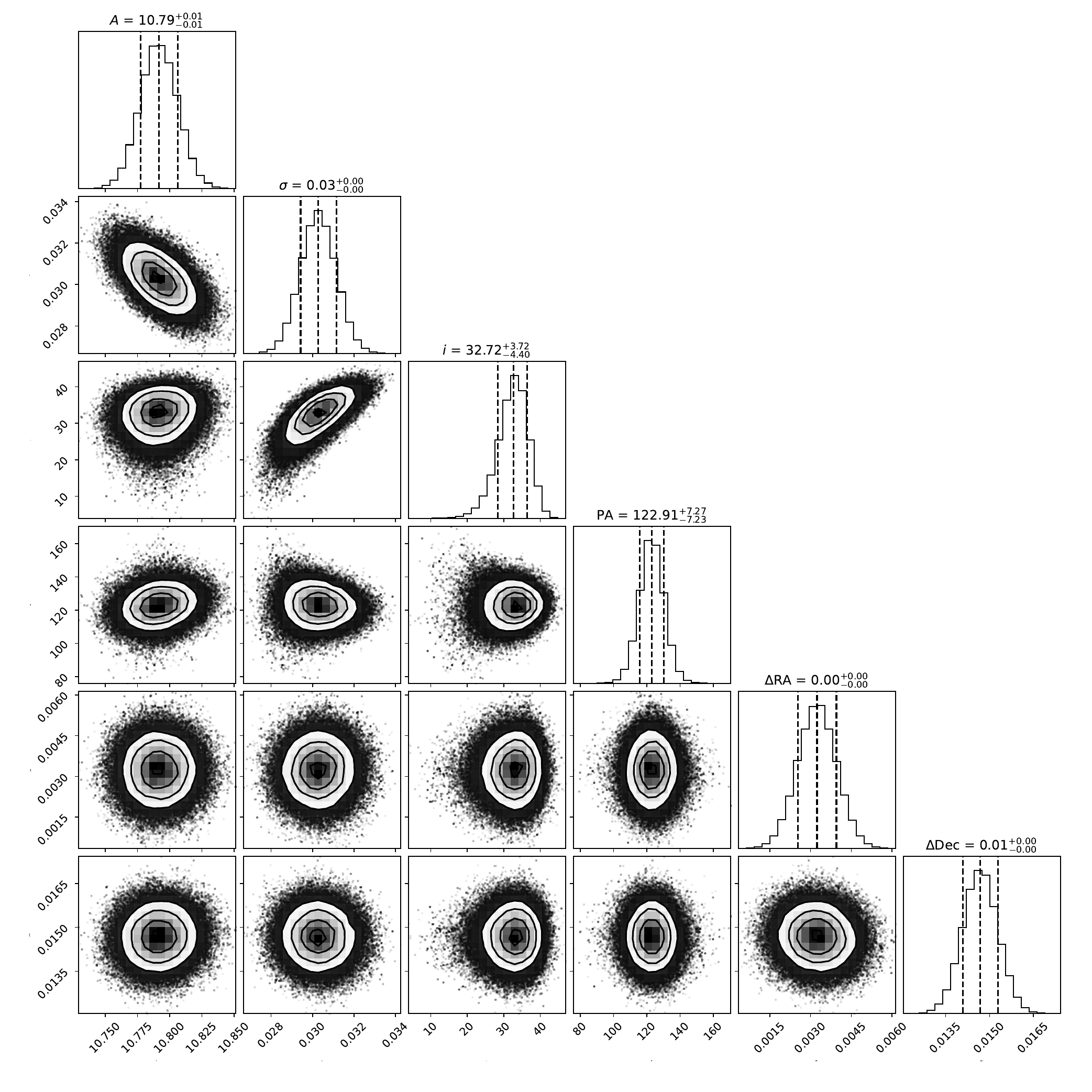}
}
\caption{Visibility modeling of the single-Gaussian continuum model. \textit{Left panel:} uvplot showing the data in gray and the best-fit in green. The residual are shown in the bottom panel. \textit{Right panel:} Corner plot showing the posterior distributions and parameter correlations for the free parameters of the best-fit model obtained from the MCMC analysis.} 
\label{galario_plots_only_gauss}
\end{figure}

\begin{figure}[htb!]
\centering
\subfigure{
   \includegraphics[width=8cm]{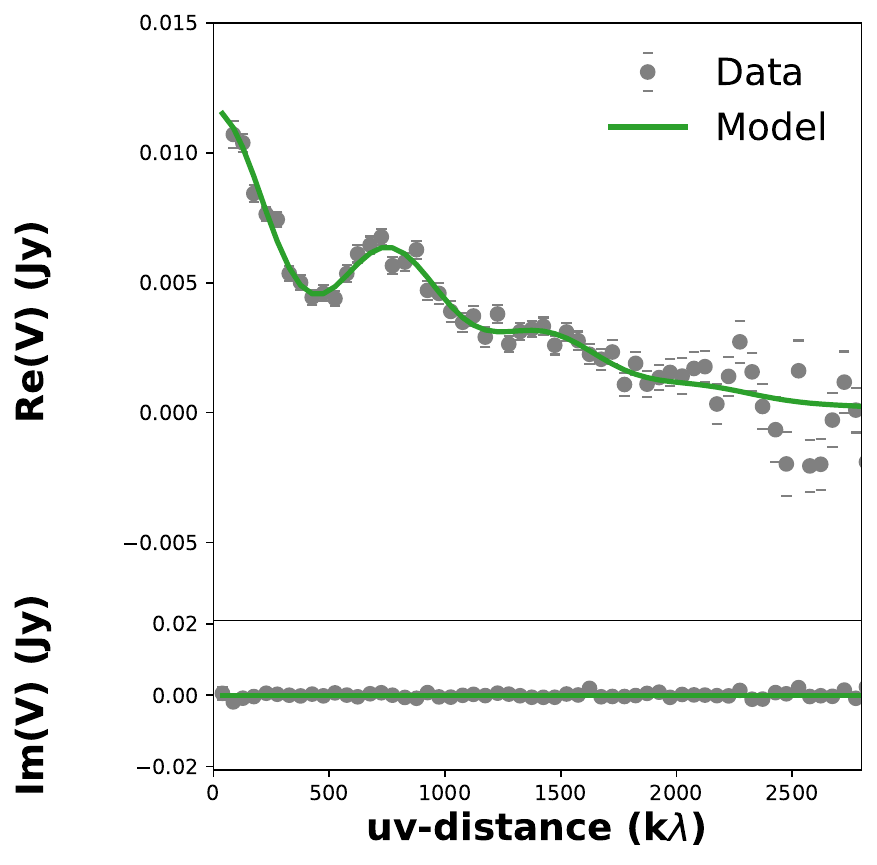}
}
\subfigure{
   \includegraphics[width=8cm]{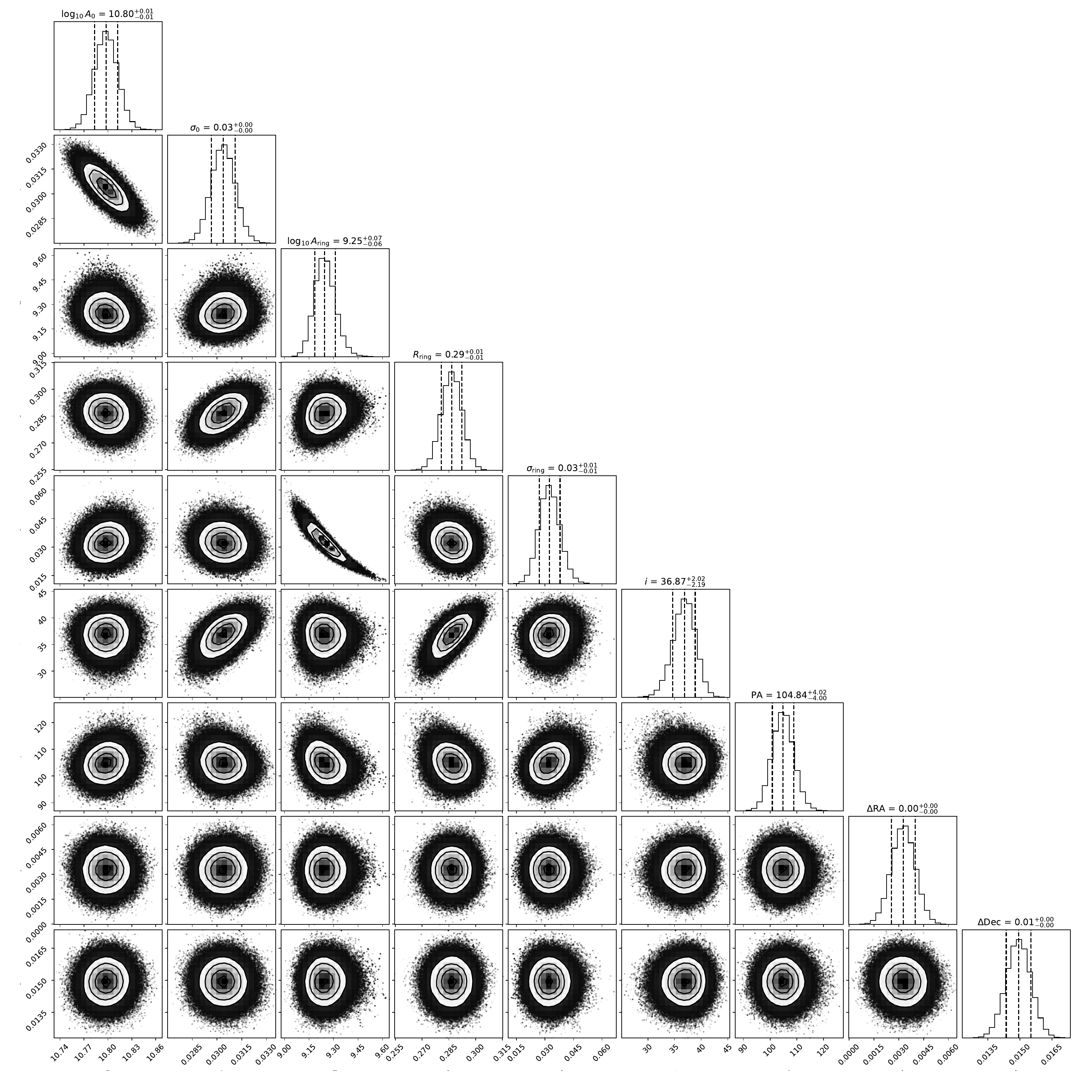}
}
\caption{Visibility modeling of the Gaussian+ring continuum model. \textit{Left panel:} uvplot showing the data in gray and the best-fit in green. The residual are shown in the bottom panel. \textit{Right panel:} Corner plot showing the posterior distributions and parameter correlations for the free parameters of the best-fit model obtained from the MCMC analysis.} 
\label{galario_plots_1ring}
\end{figure}

This appendix presents the best-fit plots by Galario for the simple Gaussian profile (Fig.~\ref{galario_plots_only_gauss}) and the Gaussian+ring profile (Fig.~\ref{galario_plots_1ring}).

\newpage
\section{Comparison with the literature} \label{comparison_lit}

\begin{figure}[htb!]
\centering
\subfigure{
   \includegraphics[width=8cm]{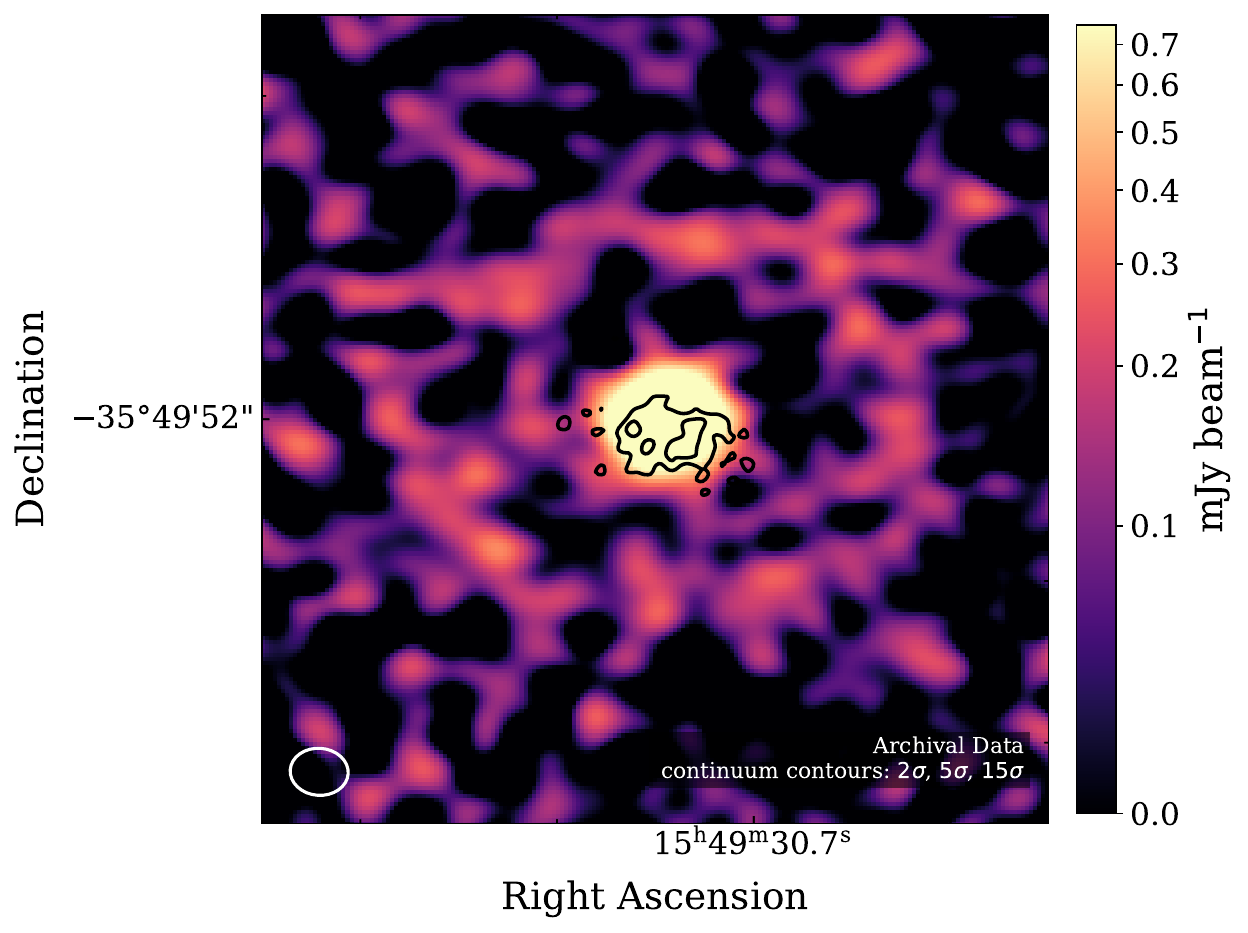}
}
\subfigure{
   \includegraphics[width=8cm]{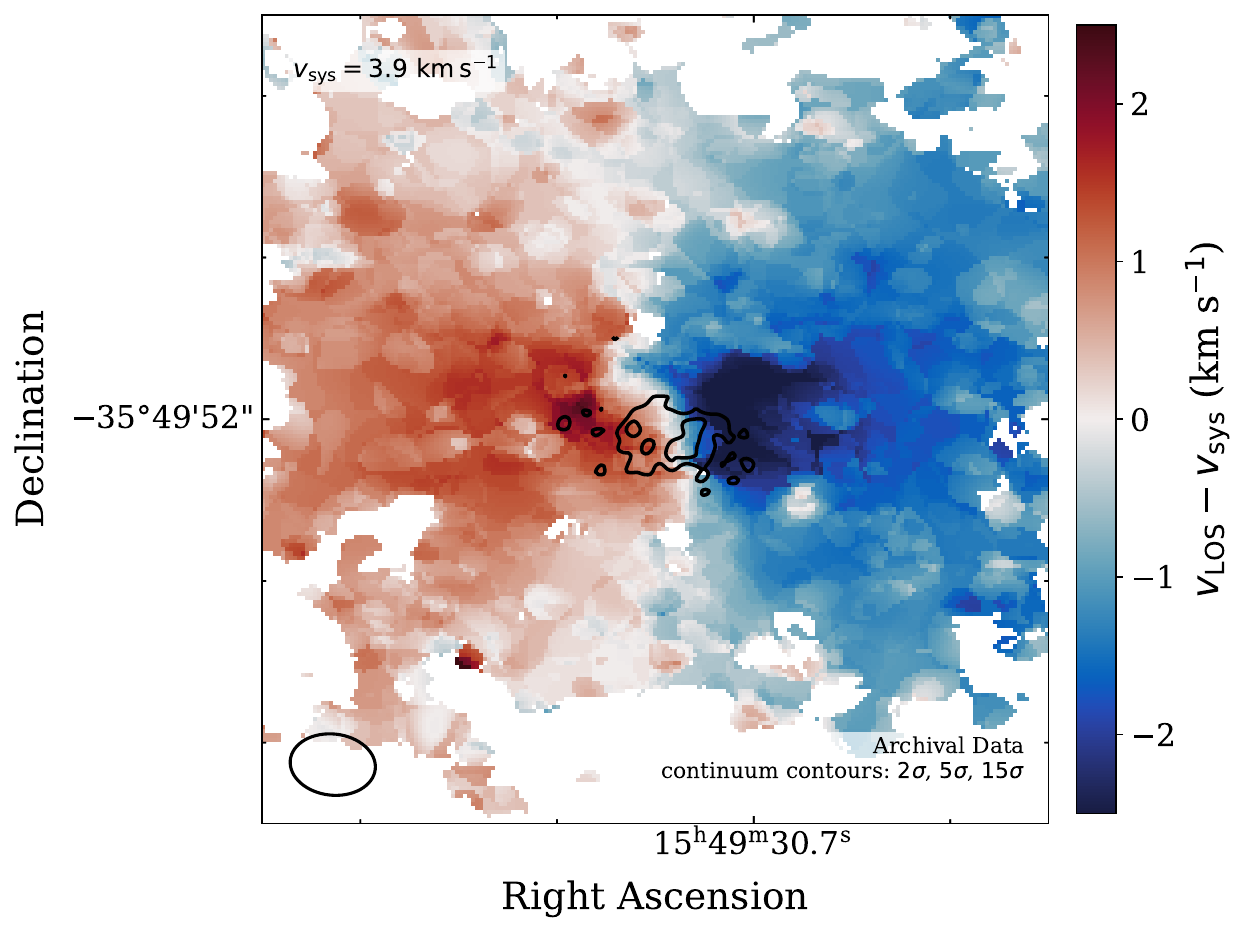}
}
\caption{\textit{Left panel:} ALMA Band 7 continuum image of Sz 76. \textit{Right panel:} Intensity-weighted velocity (moment 1) map of the $^{12}$CO ($J=3-2$) emission. In both panels, black contours show the continuum emission from the archival Band 6 observations adopted by \cite{GuerraAlvarado25}, at $[2,5,15]\sigma$. The synthesized beam is shown in the lower left corner of each panel.} 
\label{old_contours}
\end{figure}

In this appendix, we compare the data presented in this Letter with previous ALMA Band 6 observations presented by \citet{GuerraAlvarado25}. In Fig.~\ref{old_contours}, we show our continuum and $^{12}$CO moment-1 maps overlaid with the Band 6 continuum contours, after correcting the latter for the \textit{Gaia}-based proper motion of Sz 76 ($\mu_\alpha = -12.828\pm0.027$ mas/yr and $\mu_\delta = -23.703 \pm 0.022$ mas/yr). Although the $\sim$4 au depletion remains unresolved at our angular resolution, the overlay contours place this compact inner structure in our larger-scale dust and gas disk.

\section{Exploratory planet–disk interaction analysis} \label{DBnet}

\begin{figure}[htb!]
    \centering
    \includegraphics[width=\columnwidth]{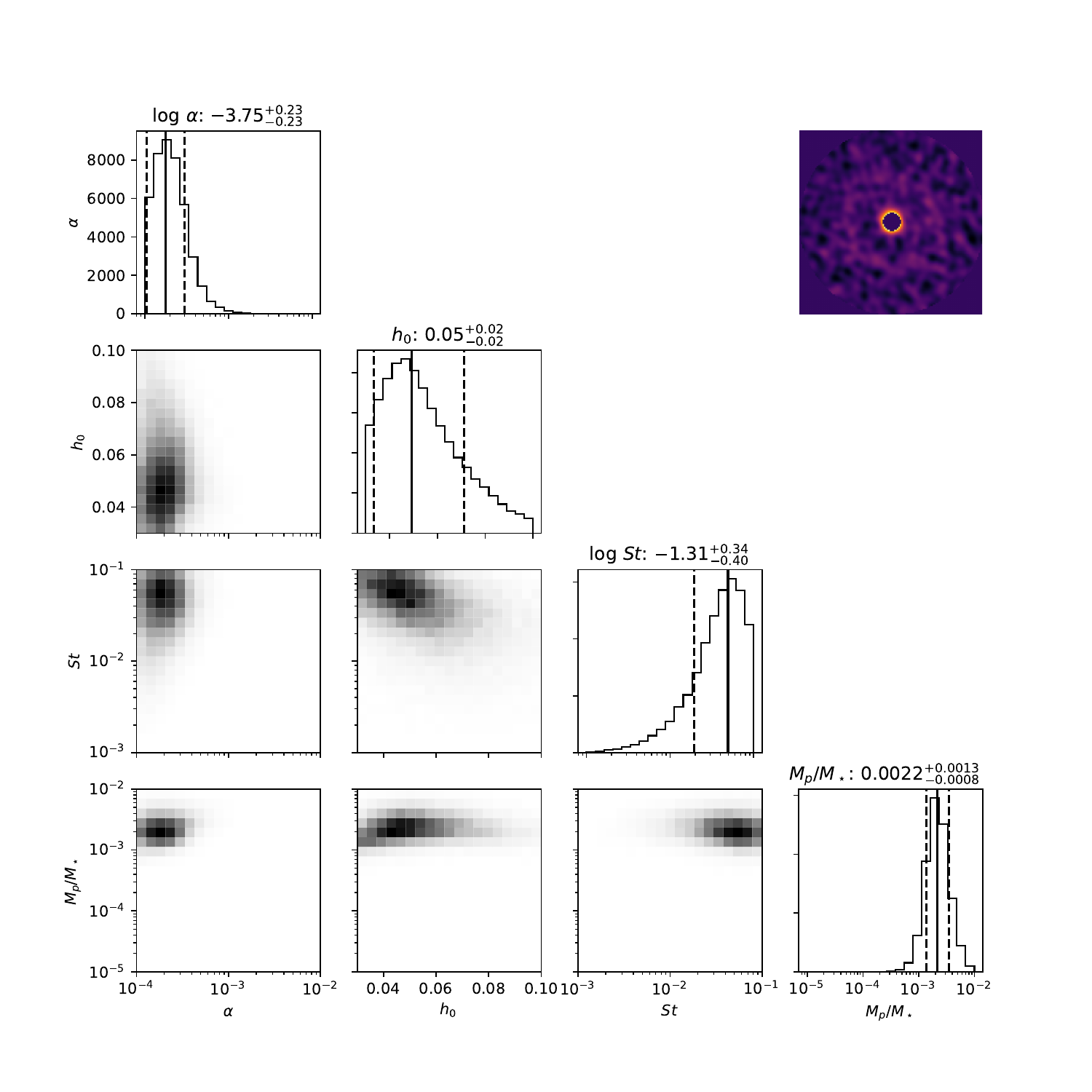}
    \vspace{-0.7cm}
    \caption{Posterior distributions of the disk and planet parameters inferred with \texttt{DBNets2.0} for Sz~76, assuming a planet location of $R_p = 23.46$au. The image in the right, upper corner plot is the deprojected disk image.} 
    \label{DBNet1}
\end{figure}

The ring-like morphology identified in the continuum emission of Sz~76 suggests that the observed dust substructure may be associated with a local pressure maximum generated by an embedded planet. To investigate this, we used \texttt{DBNets2.0} \citep{Ruzza25}, a simulation-based inference framework that infers the properties of potential planets and their host disks from dust continuum substructures. This framework simultaneously constrains the planet-to-star mass ratio, disk $\alpha$-viscosity, aspect ratio (h) and dust Stokes number ($S_t$). Using the disk geometry obtained with Galario, a distance of 156.4 pc, and $M_{\star}=0.19 M_{\odot}$, and initially placing a putative planet at $R_p = 23.46$ au (equal to the minimum of the radial brightness profile between the bright central component and the outer ring), we find a planet-to-star mass ratio of $\frac{M_p}{M_{\star}}=0.0022^{+0.0013}_{-0.0008}$, corresponding to a mass planet of $M_p = 0.44^{+0.26}_{-0.16} M_{jup}$. The inferred disk parameters are log$\alpha = -3.75 \pm 0.23$, h = 0.05 $\pm$ 0.02, and log$S_t = -1.31^{+0.34}_{-0.40}$. The resulting posterior distributions are shown in Fig.~\ref{DBNet1}.

\begin{figure}[htb!]
    \centering
    \includegraphics[width=\columnwidth]{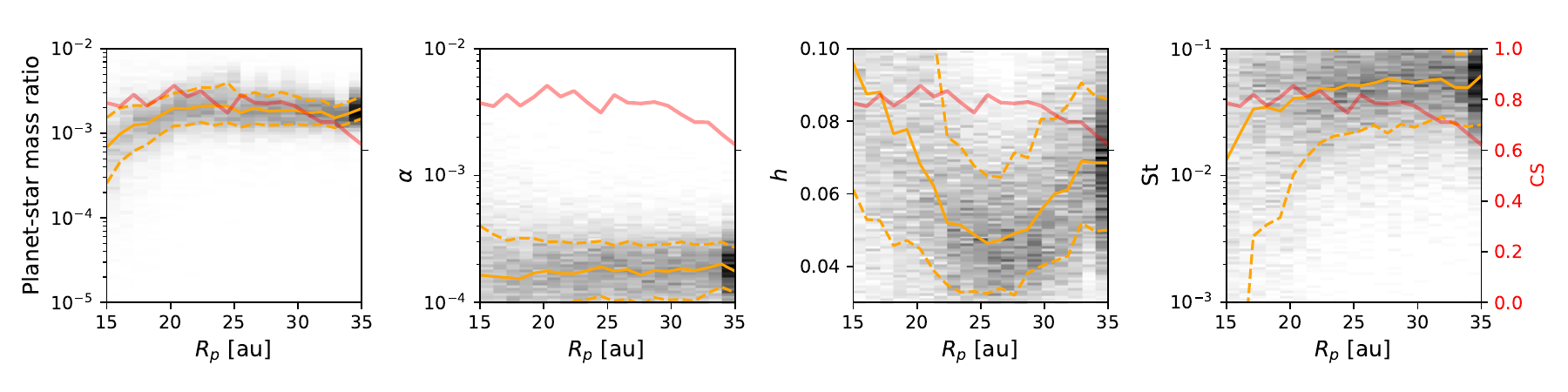}
    \vspace{-0.7cm}
    \caption{Dependence of the \texttt{DBNets2.0} inference on the assumed planet location $R_p$. The confidence score is shown as a red curve.}
    \label{DBNet2}
\end{figure}

As the location of the planet is unknown, we tested the robustness of the inference by repeating the analysis for assumed orbital radii between 15 and 35 au (Fig.~\ref{DBNet2}). The confidence score shows a broad maximum around $R_p \sim 20 - 25$ au, close to the minimum of the observed radial brightness distribution. Importantly, the inferred $\frac{M_p}{M_{\star}}$ remains approximately constant at $\sim 2 \dot 10^{-3}$ indicating that the inferred $M_p$ is only weakly dependent on the precise assumed location within the gap. The inferred $\alpha$ shows a similarly weak dependence on $R_p$, while the aspect ratio reaches its lowest values for $R_p = 25$au. 

Overall, assuming that the observed dust morphology is produced by planet-disk interaction, \texttt{DBNets2.0} suggests the presence of a sub-Jovian planet located between 20 and 25 au that could reproduce the observed substructure. While this analysis does not prove the existence of an embedded planet, it provides a quantitative estimate of the properties of a potential companion associated with a planetary interpretation of the outer dust ring.

\end{appendix}
\end{document}